\documentclass[aps,showpacs,a4paper,floatfix,twocolumn,prc,amsmath,amssymb]{revtex4-1}
\usepackage{graphicx}
\usepackage{epsfig}
\usepackage{ulem,color}
\usepackage{morefloats}
\usepackage{rotating}
\usepackage{relsize}
\usepackage{url}

\begin{document}

\title{Effect  of strong magnetic fields  on the crust-core transition
  and inner crust of neutron stars}

\author{Jianjun Fang$^1$} 
\author{Helena Pais$^1$}
\author{Sagar Pratapsi$^1$}
\author{Sidney Avancini$^{2}$}
\author{Jing Li$^{3}$}
\author{Constan\c ca Provid{\^e}ncia$^1$}

\affiliation{$^1$CFisUC, Department of Physics, University of Coimbra, 3004-516 Coimbra, Portugal.\\
$^2$Departamento de F\'{\i}sica, Universidade Federal de Santa Catarina, Florian\'opolis, SC, CP. 476, CEP 88.040-900, Brazil. \\
$^3$Department of Physics, Qufu Normal University, Qufu 273165, China.
} 

\begin{abstract}
The Vlasov equation is used to determine
the dispersion relation for the eigenmodes of magnetized nuclear and
neutral stellar  matter, taking into account the anomalous magnetic
moment of nucleons. The formalism is applied to the determination of the
dynamical spinodal section, a quantity that gives a good estimation of
the crust-core transition in neutron stars.
We study the effect of strong magnetic fields, of the order of
$10^{15}-10^{17}$ G, on the extension of the crust of
magnetized neutron stars. The  dynamical instability region
of neutron-proton-electron ($npe$) 
matter at subsaturation densities is determined within a relativistic mean field model. It is
shown that a strong magnetic field has a large
effect on the instability region, defining the crust-core transition as
a succession of stable and unstable regions  due to the opening of new
Landau levels. The effect of the anomalous magnetic moment is
non-negligible for fields larger than 10$^{15}$ G.
The complexity of the crust at the transition to the core and the
increase of the crust thickness may have direct impact on the properties
of neutrons stars related with the crust.
\end{abstract}

\maketitle

\section{Introduction}

Magnetars form a class of strongly magnetized  neutron stars  that
includes  
soft $\gamma$-ray repeaters and anomalous X-ray pulsars
\cite{duncan,usov,pacz}. These stars  have strong  surface magnetic
fields of the order of $10^{13}-10^{15}$ G \cite{sgr} and, if
isolated,  present relatively large spin periods, of the order of
2--12~s. In fact, until now, no isolated X-ray pulsar has been detected  with a spin period longer
than 12~s. This feature has recently been attributed in Ref. \cite{pons13} to the fast decay of
the magnetic field, if the  existence of a  resistive amorphous
layer at the bottom of the inner crust is confirmed. It was discussed
in Ref. \cite{pons13} that this amorphous matter, characterized by a large
impurity parameter, corresponds to the matter formed by the pasta
phases proposed in Ref. \cite{pethick83}, which result from the  competition between the long-range Coulomb repulsion and short-range nuclear attraction.
 The inner crust pasta phases have been
calculated within different formalisms, including the compressible liquid
drop model  \cite{pethick83}, classical  and quantum molecular
dynamics models
\cite{pasta}, the Thomas-Fermi
approximation within relativistic nuclear models
\cite{shen98,maruyama05,avancini08}, Hartree-Fock calculations with both
non-relativistic and relativistic models\cite{gogelein08,newton09,pais12}, see Ref. \cite{oertel16} for
a review of more recent works, including calculations of interest for
core-collapse supernova.
 A recent investigation  of the
conductivity properties of the pasta phases
has shown that topological defects 
affect the electrical conductivity of the system, originating a larger
impurity parameter \cite{schneider16}.  However, in
Ref. \cite{yakovlev15}, the author has analyzed  the  electron  transport
properties in  nuclear  pasta phases in the mantle of a magnetized neutron star, and obtained 
an enhancement of the electrical conductivity. Nevertheless, it was
stressed that further studies are necessary, since the contribution of non-spherical pasta
clusters introduce  uncertainties,  and possible impurities and
defects  in nuclear  pasta should also be considered.

Nuclear matter at subsaturation densities is characterized by a
liquid-gas phase transition \cite{mueller95}. 
Moreover, since nuclear matter is formed by two different types of
particles, protons and neutrons, mechanical and chemical instabilities
may lead both to the fragmentation of a nuclear system in heavy ion
collisions and an isospin
distillation effect \cite{colonna02,chomaz04}. The region of
instability in the isospin space characterized by the proton and
neutron densities is limited by the spinodal surface
\cite{providencia06}. This surface is defined, from a thermodynamic
perspective, as the locus where the free energy curvature goes to zero,
and, from a dynamical perspective,  as the surface where the
eigenmodes of matter go to zero.  Both surfaces coincide if
perturbations of infinity wave length are considered in the dynamical description.
Spinodal decomposition has been applied  to study the fragmentation of finite
nuclear systems
within a self-consistent quantum approach in Refs.
\cite{colonna02,chomaz04}, and it was shown that  the  liquid-gas
phase transition  of  asymmetric  systems would induce a fractional distillation of the
system.

Stellar matter in neutron stars is composed of protons, neutrons,
electrons and possibly muons at subsaturation densities. These
components are in $\beta$-equilibrium, and  leptons are necessary to
neutralize matter. As referred above, at subsaturation density, stellar
matter is essentially clusterized, possibly forming pasta phases at
the higher densities. The crust-core phase transition may be
determined from the pasta phase calculations, but it has also been shown
that a very good estimation is obtained from the thermodynamical
spinodal of proton-neutron matter, or even better, the dynamical
spinodal of proton-neutron-electron matter  \cite{avancini08,avancini10}.
In fact, the same crust-core transition density  has been obtained in
Ref. \cite{avancini08}, {when} applying a  Thomas-Fermi
description of the pasta phase and a
dynamical spinodal calculation.

Understanding the properties of the crust of neutron stars  is essential because
observations of neutron stars are directly affected by them. In
particular,  an important quantity is the fractional moment of inertia of the crust
\cite{Worley08,Lattimer00}. This quantity, as suggested in
Ref. \cite{link99}, is crucial for the interpretation of the
so-called glitches, 
sudden breaks in the regular rotation of the star. Presently, it is
still not clear whether the crust is enough to describe  the
glitches correctly, or if the core also contributes, since entrainment effects
 couple the
superfluid neutrons to the solid crust \cite{chamel13,glitch2}.

We will study in the present work the effect of a strong magnetic
field on the crust-core transition, applying a dynamical spinodal
formalism, which has shown to give a good prediction of this transition
for zero magnetic field. 
This will be carried out using the relativistic Vlasov
formalism, applied to relativistic nuclear models
\cite{nielsen91,providencia06,umodes06a}, and based on a field
theoretical formulation \cite{walecka}. The normal modes of stellar
matter will be calculated and special attention will be given to the
unstable modes. We
will only consider  longitudinal modes that propagate in the direction
of the magnetic field. The effect of the magnetic field on the
spinodal surface, the crust-core transition of $\beta$-equilibrium
matter, 
the  size of the clusters in the clusterized phase, and the fractional  moment of
inertia of the crust will be studied.

Previously, there have already been studies that analyze the effect of
the magnetic field on the thermodynamical spinodal \cite{aziz08}, and the
pasta phases in the inner crust \cite{lima13}, however, both studies
have been performed for  magnetic fields more intense than the ones expected to
exist in the crust of a magnetar. 
 The effect of the magnetic field on the outer crust  was analyzed
in Ref. \cite{chamel12}, within an  Hartree-Fock-Bogoliubov calculation, and
it was shown that the Landau quantization of the  electron motion
could affect the outer crust  equation of state, giving rise to more
massive outer crusts than the expected in usual neutron stars. Also,
the neutron drip density and pressure are affected by a strong
magnetic field, showing typical quantum oscillations, which shift the
transition outer-inner crust to larger or smaller densities
\cite{chamel15}, according to the field intensity.
The present  work aims at studying the effect of the magnetic field
on the crust-core transition, and  completes the
one in Ref. \cite{fang16}, where this study was first introduced. We
present the formalism that was not introduced in Ref. \cite{fang16}, and we
discuss the importance of including   the anomalous magnetic
field.  
A relativistic
mean-field (RMF) model, that satisfies several accepted laboratory and astrophysical
constraints \cite{fortin16,dutra14},  will be considered. This is important because, depending on
the proton fraction, which is determined by the density dependence of
the symmetry energy, the magnetic field will have a weaker or stronger effect.
We will also choose realistic proton fractions in the range of
densities of interest. 
The paper is organized as follows: in section II, the formalism is introduced, in  section III, the results
of the calculations are presented and discussed, and, finally, in the
last section, the main conclusions are drawn.

\section{Formalism}

In this work, we describe stellar matter  within the nuclear RMF formalism under the effect of strong magnetic fields
\cite{broderick,aziz08}. We also analyze the effect of the anomalous
magnetic moment (AMM) in the calculation of the dynamical spinodals. In Subsection \ref{rmf}, the Lagrangian density of the RMF model is presented and in subsection \ref{spinodals}, the Vlasov formalism is discussed in detail. 

\subsection{RMF model under strong magnetic fields}
\label{rmf}

 We consider a system of nucleons with mass $M$  that interact with and through meson fields. This system is neutralized by electrons because we also want to describe stellar matter. 
The charged particles, protons and electrons, interact through the static electromagnetic field $A^{\mu}$,
$
 A^{\mu}=(0,0,Bx,0),
$
so that $\bf B$=$B$ $\hat{z}$ and $\nabla \cdot {\bf A}=$0. We consider that the electromagnetic field is externally generated, which means that only frozen-field configurations are considered in the calculations.

The Lagrangian density of our system, with $c=\hbar=$1, reads
\begin{eqnarray}
{\cal L}=\sum_{i=p,n} {\cal L}_i + {\cal L}_e + {\cal L}_{A} + \cal L_\sigma + {\cal L}_\omega + {\cal L}_\rho + {\cal L}_{\omega\rho},
\end{eqnarray}
where ${\cal L}_i$ is the nucleon Lagrangian density, given by
\begin{eqnarray}
{\cal L}_i&=&\bar \psi_i\left[\gamma_\mu i D^\mu-M^*-\frac{1}{2}\mu_N\kappa_b\sigma_{\mu \nu} F^{\mu \nu}\right]\psi_i,
\end{eqnarray}
with
\begin{eqnarray}
M^*&=&M-g_s\phi, \\
iD^\mu&=&i \partial^\mu-g_v V^\mu-
\frac{g_\rho}{2}\boldsymbol\tau \cdot \mathbf{b}^\mu - e A^\mu
\frac{1+\tau_3}{2},
\end{eqnarray}
and the electron Lagrangian density, ${\cal L}_e$, together with the electromagnetic term, ${\cal L}_{A}$, are given by
\begin{eqnarray}
{\cal L}_e&=&\bar \psi_e\left[\gamma_\mu\left(i\partial^\mu + e A^\mu\right)-m_e\right]\psi_e, \\
{\cal L}_{A}&=&-\frac{1}{4}F_{\mu\nu}F^{\mu\nu}.
\end{eqnarray}

The electromagnetic coupling constant is given by $e=\sqrt{4\pi/137}$, and
$\tau_{3}=\pm 1$ is the
isospin projection for  protons ($+1$) and neutrons ($-1$). The nucleon AMM are introduced via the coupling of the baryons to the electromagnetic
field tensor, $F_{\mu\nu}=\partial_\mu A_\nu-\partial_\nu A_\mu$, with $\sigma_{\mu\nu}=\frac{i}{2}\left[\gamma_{\mu},
  \gamma_{\nu}\right] $, and strength $\kappa_{b}$, with
$\kappa_{n}=-1.91315$ for the neutron, and $\kappa_{p}=1.79285$ for the
proton. $\mu_N$ is the nuclear magneton. The contribution of the  anomalous magnetic moment of the
electrons is negligible \cite{duncan00}, hence it will not be considered.

We consider three meson fields, where the isoscalar part is associated with the scalar sigma ($\sigma$) field $\phi$ with mass $m_s$, and the vector omega ($\omega$) field $V^\mu$ with mass $m_v$, whereas the isospin dependence comes from the isovector-vector rho ($\rho$) field $b^i_\mu$ (where $\mu$ stands for the four-dimensional spacetime indices and $i$ is the three-dimensional isospin direction index) with mass $m_{\rho}$. The
associated Lagrangians are
\begin{eqnarray}
{\cal L}_\sigma&=&\frac{1}{2}\left(\partial_\mu\phi\partial^\mu\phi-m_s^2 \phi^2 - \frac{1}{3}\kappa \phi^3 -\frac{1}{12}\lambda\phi^4\right),\nonumber\\
{\cal L}_\omega&=&-\frac{1}{4}\Omega_{\mu\nu}\Omega^{\mu\nu}+\frac{1}{2}
m_v^2 V_\mu V^\mu, \nonumber \\
{\cal L}_\rho&=&-\frac{1}{4}\mathbf B_{\mu\nu}\cdot\mathbf B^{\mu\nu}+\frac{1}{2}
m_\rho^2 \mathbf b_\mu\cdot \mathbf b^\mu,
\end{eqnarray}
where
$\Omega_{\mu\nu}=\partial_\mu V_\nu-\partial_\nu V_\mu$, and $\mathbf B_{\mu\nu}=\partial_\mu\mathbf b_\nu-\partial_\nu \mathbf b_\mu
- g_\rho (\mathbf b_\mu \times \mathbf b_\nu)$.

The NL3$\omega\rho$ model \cite{hor01}, that we are going to consider throughout the calculations, has an additional nonlinear term, ${\cal L}_{\omega\rho}$, that mixes the $\omega$ and $\rho$ mesons, allowing to soften the density dependence of the symmetry energy above saturation density. This term is given by
\begin{eqnarray}
\mathcal{L}_{\omega \rho } &=& \Lambda_v g_v^2 g_\rho^2 V_{\mu }V^{\mu }
\mathbf{b}_{\mu }\cdot \mathbf{b}^{\mu }.
\end{eqnarray}
Some of the saturation properties of NL3$\omega\rho$ are: the binding
energy, $B=-16.2$ MeV, the saturation density,  $\rho=0.148$ fm$^{-3}$, the
incompressibility, $K= 272$ MeV, the  symmetry energy, $J=31.7$ MeV, and its slope, $L= 55.5$
MeV. The model satisfies the constraints imposed by microscopic
calculations of neutron matter \cite{neutron}, and it predicts stars with
masses above 2$M_\odot$, even when hyperonic degrees of freedom are
considered \cite{fortin16}.

\subsection{Dynamical spinodal under strong magnetic fields}
\label{spinodals}

In this subsection, we show in detail the formalism already introduced in Ref. \cite{fang16}, where the dynamical spinodals are calculated within the Vlasov formalism, as previously 
discussed in  \cite{providencia06,nielsen91,pais16V}.

The  distribution function for $npe$ matter at position $\mathbf r$,
instant $t$, and momentum $\mathbf{p}$, is given by
\begin{equation}
f({\bf r},{\bf p},t)=\mbox{diag}(f_{p},\,f_{n},\,f_{e})
\label{dis}
\end{equation}
   and 
$
h=\mbox{diag}\left(h_{p},h_{n},h_{e}\right)
$
is the corresponding one-body hamiltonian, where
\begin{eqnarray}
h_i&=& \epsilon_i + {\cal V}^i_{0},\quad  
\epsilon_i=\sqrt{(\bar{\boldsymbol{p}}^i_z)^2+\bar m_i^2}, \quad i=p,e \; \\
h_n&=& \epsilon_n + {\cal V}^n_{0}, \;
\epsilon_n=\sqrt{\left(\bar{\boldsymbol{p}}_z^n\right)^2+\left({\epsilon^n_\perp}-s\mu_N\kappa_nB\right)^2},
\end{eqnarray}
with $\bar{\boldsymbol{p}}^i=\boldsymbol{p-\mathcal{V}}^i$, and
\begin{eqnarray*}
\bar m_p&=&\sqrt{M^{* 2}+2\nu eB}-s\mu_N\kappa_p B,\\
\bar m_e&=&\sqrt{m^{* 2}_e+2\nu eB},\\
\epsilon^n_\perp&=&\sqrt{M^{*
    2}+\left( \bar{\boldsymbol{p}}_\perp^n\right)^2},\\ 
{{{\cal V}^n_\mu}}&=& g_v {V}_\mu -\frac{g_\rho}{2}\,{ b}_\mu,\\
 {\cal
  V}^p_{\mu}&=& g_v V_\mu  + \frac{g_\rho}{2}\, b_\mu+ e\, A_\mu,\\ 
 {\cal  V}^e_{\mu}&=& - e\, A_\mu.
\end{eqnarray*}
$\nu=n+\frac{1}{2}-{\rm sgn}(q)\frac{s}{2}=0, 1, 2, \ldots$ enumerates
the Landau levels of the fermions with electric charge $q$, and $s$ is the
quantum number spin, with $+1$ for spin up, and $-1$ for spin down. The vectors (${\boldsymbol{p}},\,{\boldsymbol{V}},...$) are defined along parallel ($\boldsymbol{p}_z,\,\boldsymbol{V}_z,...$) and perpendicular (${\boldsymbol{p}_{\perp}},\, {\boldsymbol{V}_{\perp}},...$) directions, since the magnetic field is taken in the $z$-direction.

The Vlasov equation is given by
\begin{equation}
\frac{\partial f_i}{\partial t} +\{f_i,h_i\}=0,\;i=p,\,n,\,e,
\label{vla}
\end{equation}
and describes the time evolution of the distribution function. 
 $\{,\}$ denotes the Poisson brackets.

The equations, describing the time evolution of the fields $\phi$,  $V^\mu$, $A^\mu$, and the third component of
the $\rho$-field $b_{3\mu}=(b_0,\mathbf{b})$, are derived from the Euler-Lagrange formalism:
\begin{equation}
\frac{\partial^2\phi}{\partial t^2} - \nabla^2\phi +m_s^2\phi +
\frac{\kappa}{2} \phi^2 + \frac{\lambda}{6} \phi^3
= g_s [\rho^p_{s}+\rho^n_{s}] \;,
\label{eqphi0}
\end{equation}
\begin{equation}
\frac{\partial^2 V_\mu}{\partial t^2} - \nabla^2 V_\mu + m_v^2 V_\mu\, +2\Lambda_v g_v^2 g_\rho^2 
{b}_{3\mu }\cdot{b}^{3\mu }V_{\mu } =\,
g_v [j^p_{\mu} +j^n_{\mu}] \;,
\label{eqv00}
\end{equation}
\begin{equation}
\frac{\partial^2 b_{3\mu}}{\partial t^2} - \nabla^2 b_{3\mu} + m_\rho^2 b_{3\mu}\, +2\Lambda_v g_v^2 g_\rho^2 V_{\mu }V^{\mu }
{b}_{3\mu }=\,
\frac{g_\rho}{2}[j^p_{\mu} -j^n_{\mu}] \;,
\label{eqb00}
\end{equation}
\begin{equation}
\frac{\partial^2 A_\mu}{\partial t^2} - \nabla^2 A_\mu \, =\,
e [j^p_{\mu} -j^e_{\mu}] \;,
\label{eqA00}
\end{equation}
where the scalar densities are given by
\begin{equation}
\rho^p_{s}({\bf r},t)=\frac{eB}{(2\pi)^2}\sum_{\nu,s}\int dp_z f_p\frac{\bar m_p M^*}{(\bar m_p+s\mu_N\kappa_p B)\epsilon_p} \;,
\nonumber
\end{equation}
\begin{equation}
\rho^n_{s}({\bf r},t)=\frac{1}{(2\pi)^3}\sum_{s}\int d^3p f_n\left(1-\frac{s\mu_N\kappa_n B}{\sqrt{M^{* 2}+p^2_\perp}}\right)\frac{M^*}{\epsilon_n}\;,
\nonumber
\end{equation}
and the components of the four-current density are
\begin{eqnarray}
j^i_{0}({\bf r},t)&=&\rho_i = \frac{eB}{(2\pi)^2}\sum_{\nu,s}\int f_i({\bf
  r},{\bf p},t)dp_z,\quad i=p,e \nonumber \;,\\
j^n_{0}({\bf r},t)&=&\rho_n = \frac{1}{(2\pi)^3}\sum_{s}\int f_n({\bf r},{\bf p},t)d^3p \nonumber\;,\\  
{\bf j}^i({\bf r},t)&=&\frac{eB}{(2\pi)^2}\sum_{\nu,s}\int f_i({\bf
  r},{\bf p},t)\frac{\bar{\boldsymbol{p}}^i_z}{\epsilon_i} dp_z, \quad i=p,e
\nonumber,\\
{\bf j}^n({\mathbf r},t)&=&\frac{1}{(2\pi)^3}\sum_{s}\int f_n({\mathbf
  r},{\mathbf p},t) d^3p \nonumber \\
&\times&\left[\frac{\bar{\boldsymbol{p}}^n_z}{\epsilon_n}
 +\frac{\bar{\boldsymbol{p}}^n_\perp}{\epsilon_n}\left( 1- \frac { s\mu_{N}\kappa_{n}B } {\sqrt{M^{* 2}+(\bar{\boldsymbol{p}}^n_\perp)^2}} \right)\right] \; ,
\end{eqnarray}

As explained in Ref. \cite{fang16}, the summation in $\nu$ in the
above expressions terminates at $\nu_{max}^i \, (i=p,e)$, which is the
largest value of $\nu$ for which the square of the Fermi momenta of
the particle is still positive, and which corresponds to the closest
integer from below, defined by the ratio 
\begin{eqnarray*}
\nu_{max}^p&=&\frac{(\epsilon^p_F+s\mu_N\kappa_pB)^2-{M^*}^2}{2eB},\\
\nu_{max}^e&=&\frac{{\epsilon^e_F}^2-m_e^2}{2eB},
\end{eqnarray*}
 where $\epsilon^{p}_{F}$ and $\epsilon^{e}_{F}$ are the Fermi energies of protons and electrons, respectively.

At zero temperature, the ground state of the system is characterized
by the Fermi momenta $P_{F}^i \, (i=p,n,e),$ and is described by the equilibrium distribution
function 
$$f_0({\bf r},{\bf p}) = \mbox{diag}[\Theta(P_F^{p2}-p^2),\Theta(P_F^{n2}-p^2),\Theta(P_F^{e2}-p^2)],$$

where
\begin{eqnarray*}
P^p_F&=&\sqrt{\epsilon^p_F-\bar m_p^2},\\
P^n_F&=&\sqrt{\gamma-\sqrt{\gamma^2-\beta}},\\
P^e_F&=&\sqrt{\epsilon^e_F-\bar m_e^2},
\end{eqnarray*}
are the Fermi momenta of protons, neutrons and electrons,
with 
\begin{eqnarray*}
  \gamma&=&\alpha+2\left(s\mu_N\kappa_nB\right)^2\left(1-x^2\right),\\
\alpha&=&\epsilon^{n2}_F-M^{* 2}-\left(s\mu_N\kappa_nB\right)^2,\\
\beta&=&\alpha^2-4\left(s\mu_N\kappa_nB\right)^2M^{* 2},
\end{eqnarray*}
{and $x=\cos\theta'$, $\theta'$ being the polar angle. 
The equilibrium state is also defined } by the constant mesonic fields, that are given by the following equations
\begin{eqnarray}
m_s^2\phi_0 + \frac{\kappa}{2} \phi_0^2 + \frac{\lambda}{6} \phi_0^3 = g_s\rho_s^{(0)}, \\
m_v^2\,V_0^{(0)} +2\Lambda_v g_v^2 g_\rho^2V_0^{(0)}b_0^{(0)2}\,=\, g_v j_0^ {(0)}, \\
m_{\rho}^2\,b_0^{(0)}+2\Lambda_v g_v^2 g_\rho^2V_0^{(0)2}b_0^{(0)}\,=\, \frac{g_\rho}{2} j_{3,0}^{(0)}, \\
V^{(0)}_i\,= b_i^{(0)}\,=\,A_0^{(0)}\,=\, A_i^{(0)}\,=\, 0,
\end{eqnarray}
where $\rho_s^{(0)}$, $j_0^{(0)}$, $j_{3,0}^{(0)}$ are the equilibrium scalar density, the nuclear density, and the isospin density, respectively. The spatial components of
${ V^\mu},\, { b^\mu}$ and ${ A^\mu}$ are zero because there are no currents in the system.

The collective modes, which are obtained considering small oscillations around the equilibrium state, are given by the solutions of the
linearized equations of motion. The deviations from equilibrium are
described by
\begin{eqnarray*}
f_i&=& f_{0i} + \delta f_i\; ,\\
 \phi&=&\phi_0 + \delta\phi\; ,\\
 V_0&=& V_0^{(0)} + \delta V_0\;, \quad V_i\,=\,\delta V_i\; ,\\ 
  b_0&=&b_0^{(0)} + \delta b_0\; ,\quad  b_i\,=\,\delta b_i\; ,\\
 A_0&=& \delta A_0\; ,A_i\,=\,\delta A_i. 
\end{eqnarray*}
The fluctuations $\delta f_i$ are written as 
\begin{equation}
\delta f_i \,=\, \{S_i,f_{0i}\}\,=\,-\{S_i,p^2\}\delta(P^{i2}_{F}-p^2) \;,
\label{dis1}
\end{equation}
where $S_i$ are the components of a  generating function defined in
$npe$ space,
$S({\bf r},{\bf p}) = \mbox{diag}(S_p, S_n, S_e).$

The linearized Vlasov equations for $\delta f_{i}$,
$$
\frac{d\delta f_{i}}{d t}+ \{\delta f_{i}, h_{0i}\}
 +\{f_{0i},\delta h_{i} \}=0,
$$
are equivalent to the following time evolution equations \cite{nielsen91}:
\begin{equation}
\frac{\partial S_i}{\partial t} + \{S_i,h_{0i}\} = \delta h_i, \; i=p,\,n,\,e,
\label{eqevo}
\end{equation}
where
\begin{eqnarray}
\delta h_p&=&-\frac{{p}_z\cdot \delta{\cal V}_z^p}{\epsilon_{static}^p}-\frac{g_s\,M^*\bar m_p}{\epsilon_{static}^p(\bar m_p+s\mu_{N}\kappa_{p}B)}\delta\phi + \delta{\cal V}_{0p}, \nonumber \\
\delta h_n&=&-\frac{{p}_z \delta{\cal V}_z^n}{\epsilon_{static}^n}-\frac{g_s\,M^*}{\epsilon_{static}^n}\left( 1- \frac { s\mu_{N}\kappa_{n}B }{\sqrt{M^{* 2}+{\bf p_\perp}^2}}\right)\delta\phi + \delta{\cal V}_{0n}, \nonumber \\
\delta h_e&=&e\left[\frac{{p_z}\delta{A_z}}{\epsilon_{static}^e}-\delta{A}_{0}\right],
\end{eqnarray}
with
\begin{eqnarray*}
\epsilon_{static}^p&=&\sqrt{p^2_z+\bar m_p^2},\\
\epsilon_{static}^e&=&\sqrt{p^2_z+\bar m_e^2},\\
\epsilon_{static}^n&=&\sqrt{p^2_z+\left(\sqrt{M^{*2}+p^2_\perp}-s\mu_N\kappa_n B \right)^2}.
\end{eqnarray*}

In the present work, only the longitudinal modes are considered, with momentum
$\boldsymbol{k}$ in the direction of the magnetic field, and a frequency $\omega$. They are described by the following  ansatz
\begin{equation}
\left(\begin{array}{c}
S_{j}({\bf r},{\bf p},t)  \\
\delta\phi  \\
\delta \zeta_0 \\ \delta \zeta_i
\end{array}  \right) =
\left(\begin{array}{c}
{\cal S}_{\omega}^j (p,{\rm cos}\theta) \\
\delta\phi_\omega \\
\delta \zeta_\omega^0\\ \delta \zeta_\omega^i
\end{array} \right) {\rm e}^{i(\omega t - {\bf k}_z\cdot
{\bf r})} \;  ,
\label{ans}
\end{equation}
where $j=p,\, n\,, e$, $\zeta=V,\, b, A$ represent the vector  fields, and $\theta$ is the angle between ${\bf p}$ and ${\bf k}_z$.

 For these modes, we get $\delta V_\omega^x = \delta V_\omega^y=0$, $\delta b_\omega^x = \delta
b_\omega^y=0$ and $\delta A_\omega^x = \delta A_\omega^y=0$. Calling
 $\delta V_\omega^z = \delta V_\omega$, $\delta b_\omega^z = \delta
b_\omega$ and $\delta A_\omega^z = \delta A_\omega$, we have
$ \delta {\cal V}_{i,z}= \delta {\cal V}_\omega^i{\rm e}^{i(\omega t - {\bf k}_z\cdot
{\bf r})}, \qquad  \delta {\cal V}_{0i}= \delta {\cal V}_\omega^{0i}{\rm e}^{i(\omega t - {\bf k}_z\cdot
{\bf r})}.$
\begin{widetext}
Replacing the ansatz (\ref{ans}) in Eqs. (\ref{eqevo}), we get
\begin{eqnarray}
i(\omega - \omega_{0p}\xi){\cal S}_\omega^p(\xi)&=&-g_s \frac{M^*}{\epsilon_{F}^p}\left(\frac{\bar m_p}{\bar m_p+s\mu_{N}\kappa_{p}B}\right)
\delta\phi_\omega  - V_F^p\xi \delta{\cal V}_\omega^p+ \delta {\cal V}_\omega^{0p}\;  ,
\label{eqp} \\
i(\omega - \omega_{0n}x){\cal S}_\omega^n(x)&=&-g_s \frac{M^*}{\epsilon_F^n}\left(1-\frac{s\mu_N\kappa_n B}{\sqrt{M^{* 2}+P_F^{n2}(1-x^2)}}\right)
\delta\phi_\omega - V_F^n x \delta{\cal V}_\omega^n + \delta {\cal V}_\omega^{0n} \;  ,
\label{eqn} \\
i(\omega - \omega_{0e}\xi){\cal S}_\omega^e(\xi)&=&e\left(V_F^e\xi \delta A_\omega- \delta A_\omega^{0}\right)\;  ,
\label{eqe} \\
\left(\omega^2-k^2-m^2_{s,eff}\right)\delta\phi_\omega&=&-\frac{ig_sM^*}{(2\pi)^2}\left[\sum_{\nu,s,\xi}\frac{eBk_z}{\epsilon_{F}^p}\frac{\bar
    m_p\xi{\cal S}_\omega^p(\xi)}{(\bar
    m_p+s\mu_{N}\kappa_{p}B)}\right. \nonumber\\ & & \left.+P_F^{n}\omega_{0n}\sum_{s} \int_{-1}^1  x{\cal S}^n_\omega(x)\left(1-\frac{s\mu_N\kappa_n B}{\sqrt{M^{* 2}+P_F^{n2}(1-x^2)}}\right)dx\right] 
\label{eqphi} \\
\left( \omega^2-k^2-m^2_{v,eff}\right)\delta V_\omega^0&=&\chi\delta b_\omega^0-\frac{ig_v}{(2\pi)^2}\left(\sum_{\nu,s,\xi}eBk_z\xi{\cal S}_\omega^p(\xi)+P_F^n\epsilon_F^n\omega_{on} \int_{-1}^1  x {\cal S}^n_\omega(x)dx\right) \; ,
\label{eqv1} \\
\left( \omega^2-k^2-m^2_{\rho,eff}\right)\delta b_\omega^0&=&\chi\delta V_\omega^0-\frac{ig_\rho}{2(2\pi)^2}\left(\sum_{\nu,s,\xi}eBk_z\xi{\cal S}_\omega^p(\xi)-P_F^n\epsilon_F^n\omega_{on}\int_{-1}^1 \, x {\cal S}^n_\omega(x)dx\right)\; ,
\label{eqb1} \\
\left(\omega^2-k^2\right)\delta A_\omega^0&=&-\frac{e^2B}{(2\pi)^2}ik_z\sum_{\nu,s,\xi}\xi({\cal S}_\omega^p(\xi)-{\cal S}_\omega^e(\xi))\; ,
\label{eqA1}
\end{eqnarray}
\end{widetext}
where
\begin{eqnarray*}
\omega_{0i}&=&k_zV_F^i=k_zP_F^i/\epsilon_F^i,\,i=p,n,e,\\
m^2_{s,eff}&=&m_s^2+\kappa\phi_0+\frac{\lambda}{2}\phi_0^2
+g_s^2\frac{d\rho_s}{dM^*},\\
m^2_{v,eff}&=&m_v^2+2\Lambda_vg_v^2g_\rho^2b_0^{(0)2} \\
m^2_{\rho,eff}&=&m_\rho^2+2\Lambda_vg_v^2g_\rho^2V_0^{(0)2}
\end{eqnarray*}
with $\xi=\pm 1, x={\rm cos}\theta$, and $\chi=4\Lambda_vg_v^2g_\rho^2V_0^{(0)}b_0^{(0)}$. 
From the continuity equation for the density currents, we get for the components of the vector fields

\begin{eqnarray}
k_z\,\delta V_\omega&=&\omega\delta V_\omega^0 -\frac{\omega}{\omega_v^2}\chi\delta b_\omega^0 \\
k_z\,\delta b_\omega&=&\omega\delta b_\omega^0 -\frac{\omega}{\omega_\rho^2}\chi\delta V_\omega^0 \\
k_z\,\delta A_\omega&=&\omega\delta A_\omega^0 
\label{rel1}
\end{eqnarray}
with $\omega_v^2=\omega^2-k^2-m^2_{v,eff}$ and $\omega_\rho^2=\omega^2-k^2-m^2_{\rho,eff}$.

Substituting the set of equations (\ref{eqphi})-(\ref{eqA1})  into Eqs. (\ref{eqp})-(\ref{eqe}), we get a set of five independent equations of motion in terms of the amplitudes of the proton
and neutron scalar density fluctuations, $A^{ps}_{\omega,\nu,s}$, $A^{ns}_{\omega,s}$, respectively, and in terms of the amplitudes of the proton, neutron and electron vector density fluctuations, $A^{p}_{\omega,\nu,s}$, $A^{n}_{\omega,s}$, $A^{e}_{\omega,\nu,s}$, respectively. These equations are given by
\begin{equation}
\left(\begin{array}{ccccc}
a_{11}&a_{12}&a_{13}&a_{14}&a_{15}\\
a_{21}&a_{22}&a_{23}&a_{24}&a_{25}\\
a_{31}&a_{32}&a_{33}&a_{34}&0\\
a_{41}&a_{42}&a_{43}&a_{44}&0\\
0&a_{52}&0&0&a_{55}\\
\end{array}\right)
\left(\begin{array}{c}
\sum_{\nu,s}A^{ps}_{\omega,\nu,s}\\ \sum_{\nu,s}A^p_{\omega,\nu,s}\\
\sum_{s}A^{ns}_{\omega,s}\\ \sum_{s}A^n_{\omega,s}\\ \sum_{\nu,s}A^e_{\omega,\nu,s}
\end{array}\right)
=0.
\label{det}
\end{equation}

The eigenmodes $\omega$ of the system correspond to the solutions of the dispersion relation obtained by the equations written above. The coefficients $a_{ij}$ and the amplitudes are given in the Appendix. The density fluctuations can be written as
\begin{equation}
\delta\rho_n/\delta\rho_p=\frac{A_{\omega, s}^{n}}{eBA_{\omega,\nu, s}^{p}},
\label{fluc1}
\end{equation}
\begin{equation}
\delta\rho_e/\delta\rho_p=\frac{A_{\omega,\nu, s}^{e}}{A_{\omega,\nu, s}^{p}}.
\label{fluc2}
\end{equation}

At low densities, the system has unstable modes, that are characterized by
an imaginary frequency, $\omega$. 
The dynamical spinodal surface in the ($\rho_p,\rho_n$) space, for a given wave vector $\mathbf
k$, is obtained by imposing $\omega=0$.
Inside this unstable region, we also calculate the mode with
the largest growth rate,  $\Gamma$, defined as  $\omega=i\Gamma$. This mode is the one responsible for the formation of instabilities. By taking its half-wavelength, we can get a good estimation of the size of the clusters (liquid) that appear in the mixed (liquid-gas) phase, i.e. in the inner crust of the stars \cite{umodes06a}.

\section{Numerical results and discussions}

\begin{figure}[!htbp]
\includegraphics[width=0.55\textwidth]{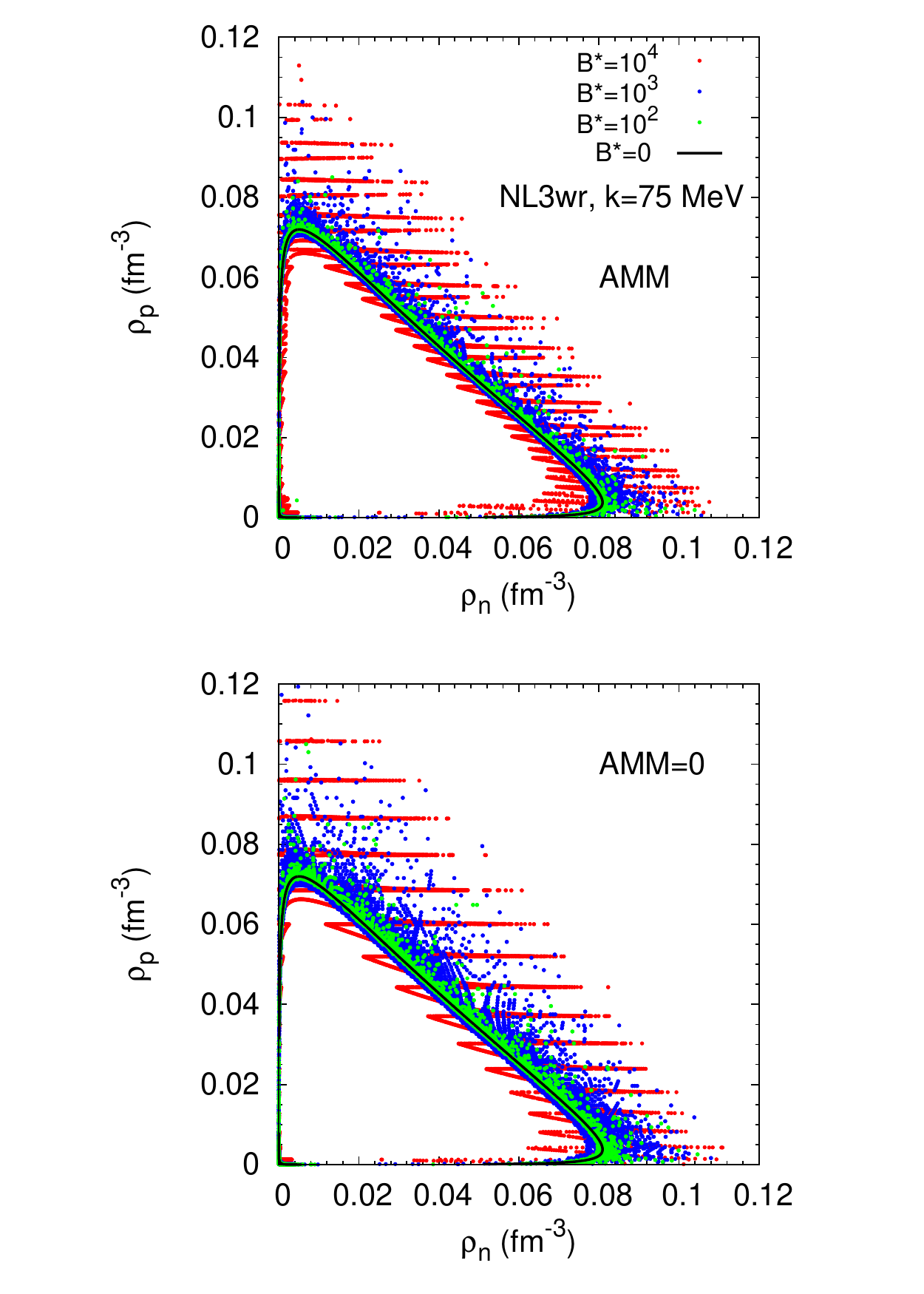} \\
\caption{Dynamical spinodal for NL3$\omega\rho$, a momentum
  transfer of $k=75$ MeV, and $B^*=10^4$ (red), $B^*=10^3$ (blue), and $B^*=10^2$ (green) with AMM (top)
  and without AMM (bottom). A comparison with the $B=0$ (black lines) results is also made.}
\label{fig1}
\end{figure}

In the present section, we discuss the effects of strong magnetic fields on the structure of 
the inner  crust of magnetars. In particular, we analyse the dynamical
spinodals for the NL3$\omega\rho$ model for three different values of
the magnetic field: $B=4.41 \times 10^{15}$ G, $B=4.41 \times 10^{16}$
G,  and $B=4.41\times 10^{17}$ G. These values correspond to
$B^*=10^{2}$, $B^*=10^{3}$, and $B^*=10^{4}$, where $B=B^*B_{ce}$, with $B_{ce}=4.41\times10^{13}$ G, being the electron critical magnetic field. In fact, the most intense fields detected on the surface of
a magnetar are not larger than $2\times 10^{15}$ G, i.e. one or two
orders of magnitude smaller than the two more intense fields 
considered in this study. However, in Refs. \cite{kiuchi08,rezzolla12}, the authors obtained toroidal fields more intense than 10$^{17}$G in stable configurations, meaning that in the interior of the stars, stronger fields may be expected.

\begin{figure}[!htbp]
\includegraphics[width=0.55\textwidth]{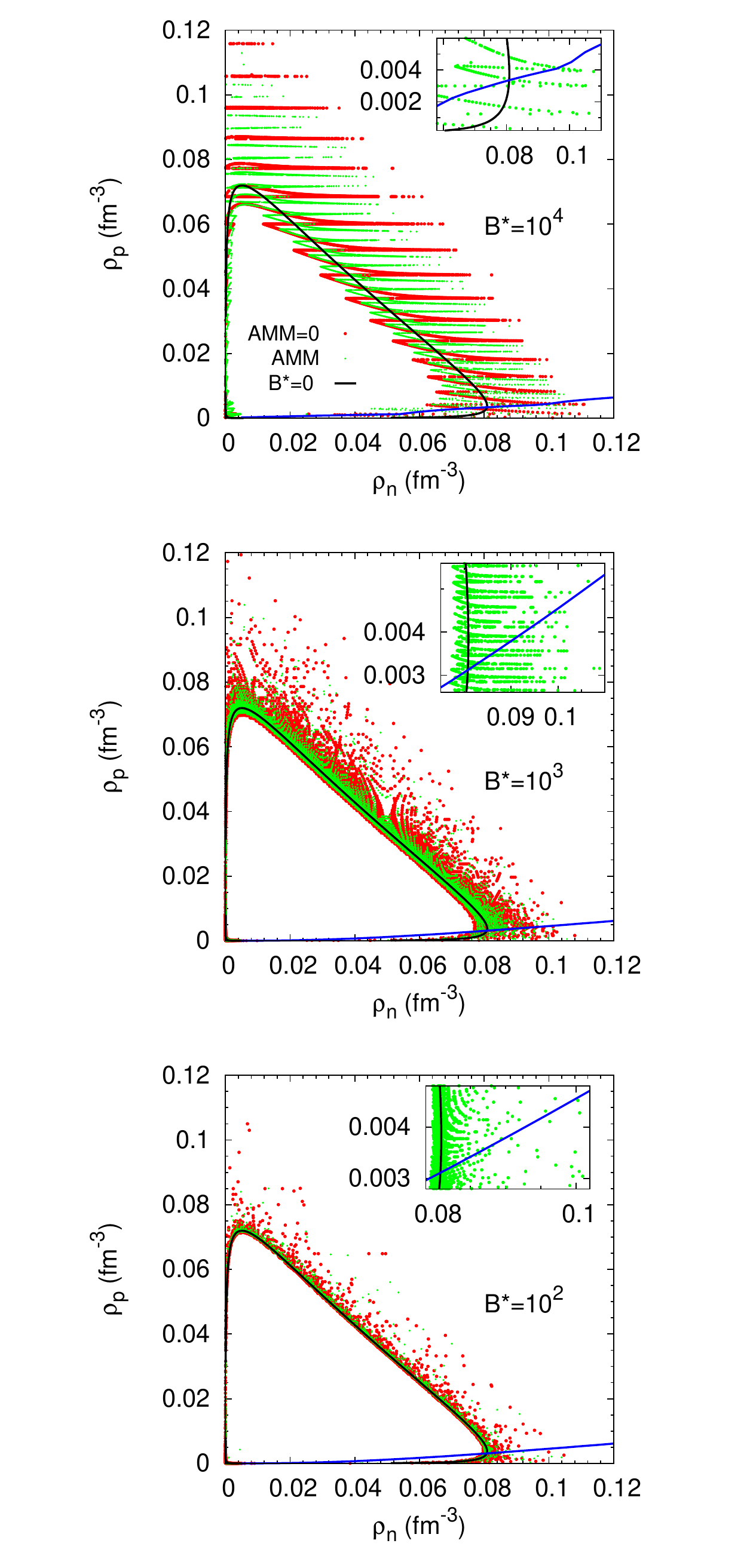} \\
\caption{Dynamical spinodal for NL3$\omega\rho$, a momentum
  transfer of $k=75$ MeV, with AMM (green) and without AMM (red), for $B^*=10^4$ (top), $B^*=10^3$ (middle), and $B^*=10^2$ (bottom). A comparison with the $B=0$ (black lines) results is also made. The EoS for $\beta-$equilibrium matter is also shown.}
\label{fig2}
\end{figure}

In Fig. \ref{fig1}, we show the dynamical spinodal sections in the
($\rho_p$,$\rho_n$) space for the  magnetic fields mentioned above,
with (top) and without (bottom panels) AMM.  The black lines
represent the spinodal section when the magnetic field is zero. The
calculations were performed with $k=75$ MeV, which is a value of the
transferred momentum that gives a spinodal section very close to
  the envelope of the spinodal sections.  
These sections have been obtained
numerically by solving the dispersion relation (\ref{det}) for
$\omega=0$. This was performed looking for the solutions at a fixed
proton fraction, and for each solution, a point was obtained. The
solutions form a large connected region for the lower proton and
neutron densities, plus extra disconnected domains that do not occur
at $B=0$. The point-like appearance of
the sections is a numerical limitation. {A higher resolution in ($\rho_p$,$\rho_n$) would complete the gaps.}

First we compare the results obtained omitting the AMM contribution
(bottom panel). The structure of the spinodal section obtained for the
strongest field considered, $B^*=10^4$, clearly shows the effect
of the Landau quantization, as already shown in Ref. \cite{fang16}: there are
instability  regions that extend to much larger densities than the
$B=0$ spinodal section, while there are also stable regions that at
$B=0$ would be unstable. This is due to the fact that the energy
density becomes softer, just after the opening of a new Landau level, and
harder when the Landau level is most filled. The spinodal section has
a large connected section at the lower densities and extra
disconnected regions. If smaller fields are considered, the structure
found for $B^*=10^4$ is still present, but at a much smaller scale due to
the increase of the number of Landau levels, see  detail in the
inset of the
middle panel of  Fig. \ref{fig2}, for $B^*=10^3$. It is clear that the
spinodal section tends to the $B=0$ one, as the magnetic field intensity
is reduced.

In the top panel of Fig. \ref{fig1}, we show the same three spinodal
sections, but with the inclusion of the AMM for the protons and neutrons. The
overall conclusions taken for the spinodals without the AMM are still
valid, although the section acquires more structure when the AMM is included
since, for each Landau level, the proton spin up and spin down levels have
different energies. This difference originates a doubling of the bands, which are
easily identified for  $B^*=10^4$. Besides, these bands are also
affected by the neutron AMM. The spinodal  sections obtained with
AMM are  smaller, as it is seen from Fig. \ref{fig2}, where, for each field intensity,
$B^*=10^4$ (top), $10^3$ (middle), and $10^2$ (bottom),
the spinodal section without (red) and with (green) AMM  are
plotted. Although the inclusion of the AMM does not have a very strong effect because the proton and neutron
anomalous magnetic moments are small, these effects are not negligible, and, in fact, they 
reduce the instability sections. In the three panels of
Fig. \ref{fig2}, we include an inset panel where we
have zoomed in the spinodal with AMM in a small range of densities  to
show that, although in a smaller scale, the structure is similar to
the one shown for $B^*=10^4$.

For neutron-rich matter, as the one occurring in neutron stars, the
instability regions extend to densities almost 40\% larger than the
crust-core transition density for $B=0$. The effect of the magnetic
field is larger precisely when the proton fraction is smaller. 
We have included in the three panels of Fig. \ref{fig2} a curve that
represents the densities ($\rho_p, \rho_n$) at $\beta$-equilibrium,
including the contribution of the AMM. The curves cross an alternation of stable and
unstable regions, indicating  the existence of a complex crust-core transition, see the insets for detail.  The beginning of an homogeneous matter is shifted to larger densities, 0.100 fm$^{-3}$ for
$B^*=10^4$, 0.103 fm$^{-3}$ for $B^*=10^3$, and 0.105 fm$^{-3}$
for $B^*=10^2$, corresponding to the pressures 0.818 MeV$/$~fm$^{3}$,
0.833 MeV$/$~fm$^{3}$, and 0.863 MeV$/$~fm$^{3}$, respectively. 
This complex transition region with a thickness of $\sim
0.02$~fm$^{-3}$, even for the weaker fields, will have strong
implications in the structure of the inner crust of magnetars. \\

 We will discuss later in more detail  the crust-core transition region in the presence of  magnetic
fields.

\begin{figure}[!htbp]
\includegraphics[width=0.5\textwidth]{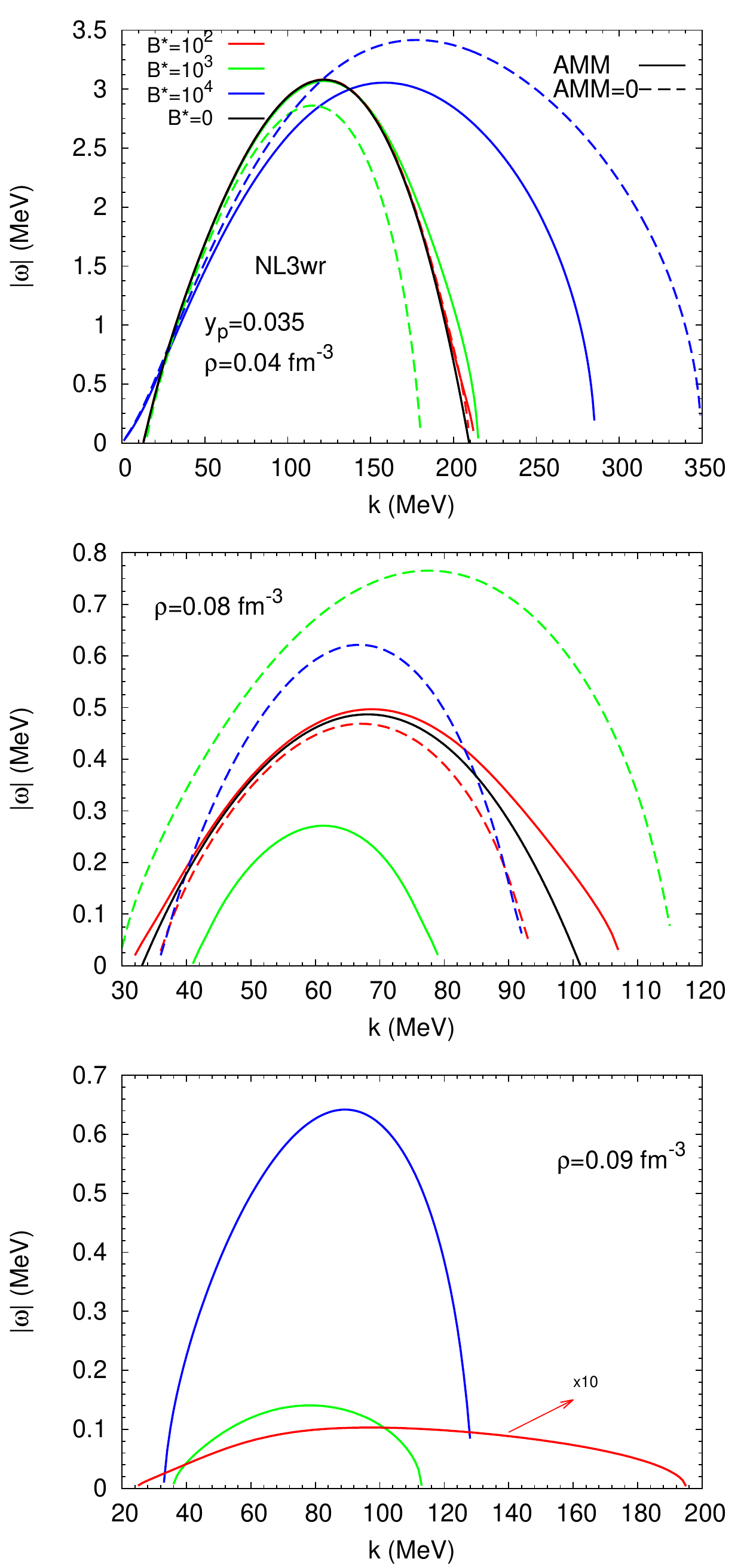} \\
\caption{Growth rates, $|\omega|$, as a function of the momentum, $k$, for NL3$\omega\rho$, a proton fraction of $y_p=0.035$, and $B^*=10^2$ (red), $B^*=10^3$ (black), and $B^*=10^4$ (blue) with AMM (solid)
  and without AMM (dashed lines), for a fixed baryonic density of
  $\rho=0.04$~fm$^{-3}$ (top), $\rho=0.08$~fm$^{-3}$ (middle), and
  $\rho=0.09$~fm$^{-3}$ (bottom). The growth rates for
    $B^*=10^2$ in the bottom panel are multiplied by a factor of 10
    and are obtained for $\rho=0.0903$~fm$^{-3}$.}
\label{fig3}
\end{figure}

The solution of the dispersion relation inside the spinodal section
gives pure imaginary frequencies, indicating that the system is unstable to
the propagation of a perturbation with the corresponding wave number
in the  density range where this occurs. The modulus
of the frequency, designated as  growth rate,  indicates how the
system evolves into a two-phase configuration. The evolution will be
dictated by the largest growth rate  \cite{chomaz04,providencia06}. 

As an example, in Fig. \ref{fig3}, we show the growth rates, $|\omega|$, as a
function of the transferred momentum $k$, for fixed values of the
baryonic density: $\rho=0.04$~fm$^{3}$ (top), $\rho=0.08$~fm$^{3}$
(middle), and $\rho=0.09$~fm$^{3}$ (bottom panels). We consider a
fixed proton fraction of 0.035, which is an average  value found for
NL3$\omega\rho$ within a Thomas-Fermi calculation of the inner crust \cite{grill12}, and we choose
the same values for the magnetic field as in the previous figures. 
The growth rates with (solid) and without (dashed) AMM are plotted
together with the growth rate at $B=0$ (black line).

We first consider $\rho=0.04$~fm$^{-3}$, far from the transition to homogeneous
matter. The smaller the field, the smaller the effect of the AMM, and
for $B^*=10^2$, the two curves superimpose, and are almost coincident
with the $B=0$ result. The effect of the AMM for
the two larger fields is non-negligible, and may go in opposite
direction because its behavior is closely related with the filling of the Landau
levels. The instability does not exist for the two smaller fields
at $k$ close to zero. This is the behavior discussed in
Ref. \cite{providencia06} and is directly related to the $1/k^2$
divergence of the Coulomb field. However, for $B^*=10^4$,  and since the electron
and proton densities are small, the
attractive nuclear interaction is strong enough to drag the electrons,
keeping the instability until $k=0$.  This is not anymore the case for
the two larger densities considered, because in these two cases, the
nuclear interaction is not able to compensate for the
larger densities of charged particles.
 The stronger nuclear attraction for $B^*=10^4$ 
is also observed for the large values of $k$: the instability is
still present for $k>300$ MeV, well above the maximum $k$ attained for
$B=0$, indicating that  the attractiveness of the nuclear force is stronger at short ranges.

 The two larger densities have been chosen because they are at the $B=0$ crust-core transition or above,
and this is the most sensitive region to the presence of a strong
magnetic field. Due to the alternation between stable and unstable
regions, it is highly probable that for one of the field intensities,
no instability is present for the particular density value
considered. This explains the non appearance of the curve with AMM for
$B^*=10^4$ and $\rho=0.08$~fm$^{-3}$. It also explains why the
behaviors with and without AMM are so different for $B^*=10^3$: the
value of the density considered picks up the instability region more or
less close to the limit of the instability region. In this case, also the maximum growth rates occur
for different wave numbers. For $B^* =10^2$, the results with and without  AMM
differ, and are not anymore coincident with the $B=0$ result, as seen for $\rho=0.04$~fm$^{-3}$. However, the maximum growth rate occurs at similar wave numbers in the three cases.


\begin{figure*}[!htbp]
\begin{tabular}{cc}
\includegraphics[width=0.5\textwidth]{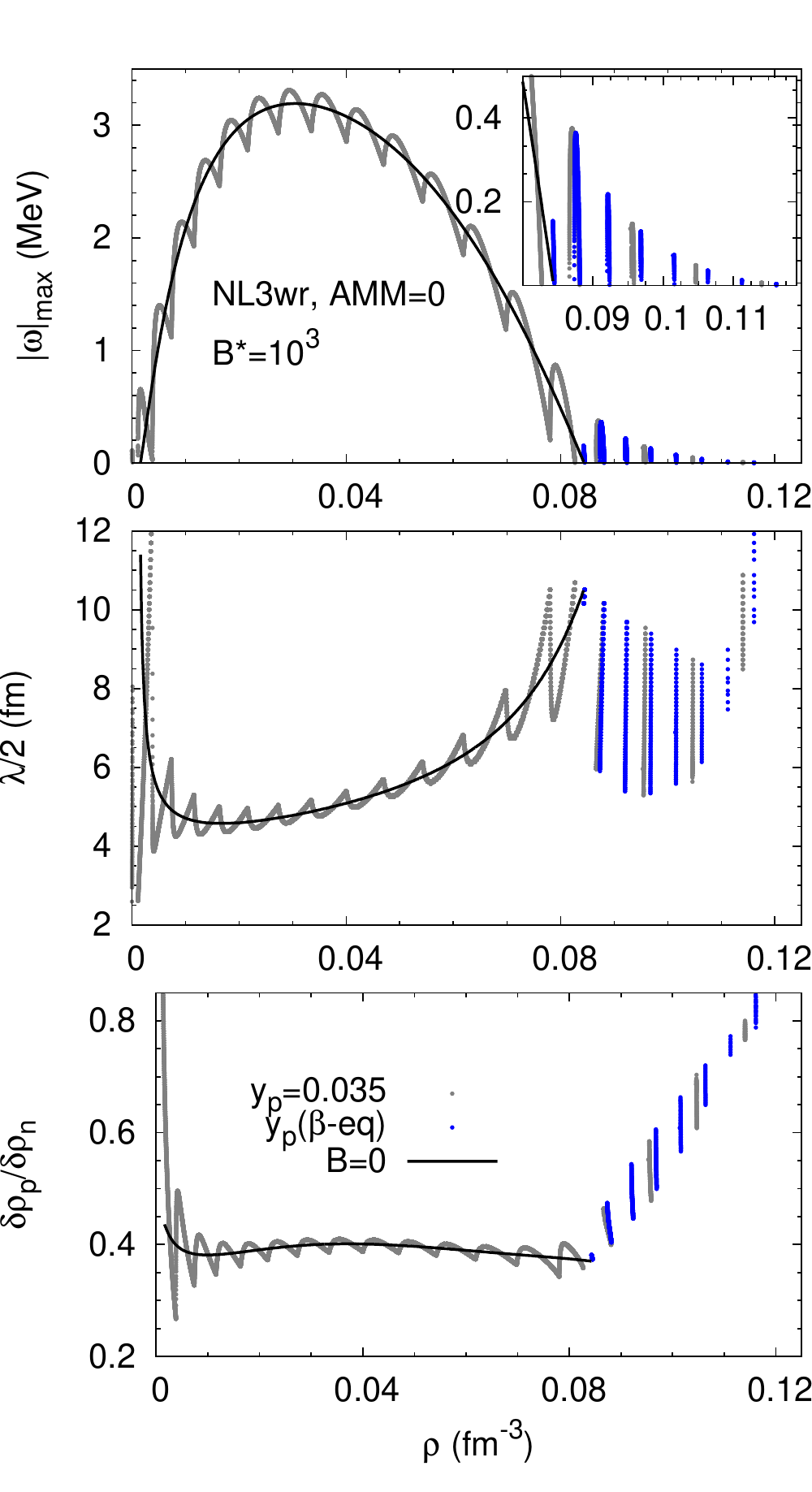} &
\includegraphics[width=0.5\textwidth]{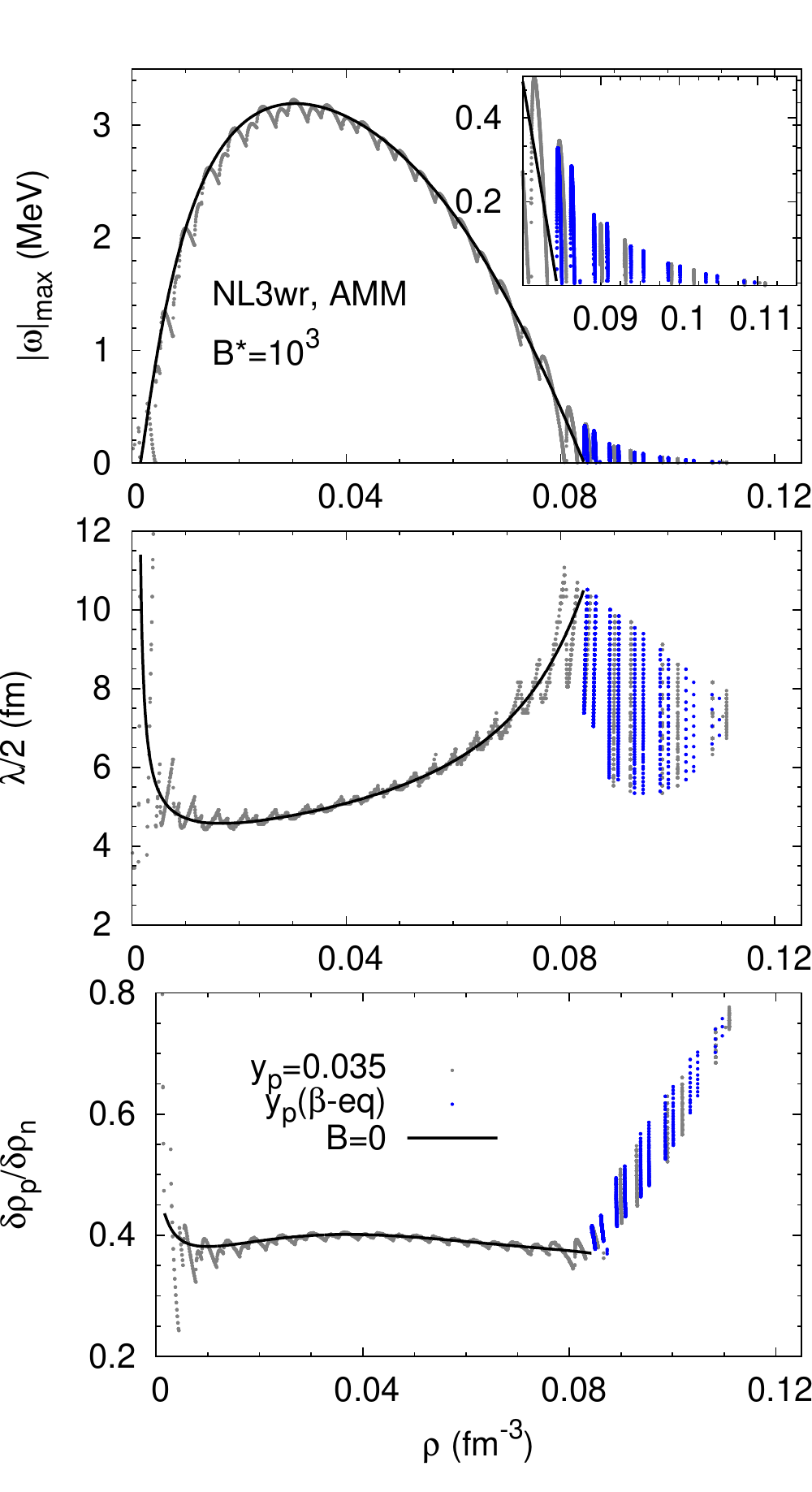} \\
\end{tabular}
\caption{Largest growth rate (top panels), the corresponding
  half-wavelength (middle panels), and the proton-neutron density
  fluctuation ratio (bottom panels) versus density for NL3$\omega\rho$,
  and a magnetic field of  $B^*=10^3$, without (left panels) and with
  (right panels) AMM. Results with the $B=0$ (black solid lines)
  calculation are also shown for comparison. The blue points
  correspond to a calculation with a proton fraction equal to the one
  found in $\beta-$equilibrium matter, above $\rho_t=0.0843$
    fm$^{-3}$, the $B=0$ crust-core transition density. The gray points correspond to a fixed proton fraction of 0.035 in the whole density range. }
 \label{fig4}
 \end{figure*}

 Finally, we consider the larger density, $\rho\sim 0.09$~fm$^{-3}$, approximately 10\% above 
the crust-core transition density, when no field is considered. 
For $B=0$, this density belongs to the core, and corresponds to
homogeneous matter. However, for the three intensities of the magnetic
field we have been considering, $B^*=10^2,\, 10^3, \, 10^4$, this is
inside or close to a region of instability. For $B^*=10^2$, we have taken
$\rho=0.0903$~fm$^{-3}$, and multiplied the growth rate by a factor of 10
in the figure. We conclude that the growth rates decrease with
the magnetic field, showing a convergence for the $B=0$ result when no
instability exists.

\begin{figure*}[!htbp]
\begin{tabular}{cc}
\includegraphics[width=0.5\textwidth]{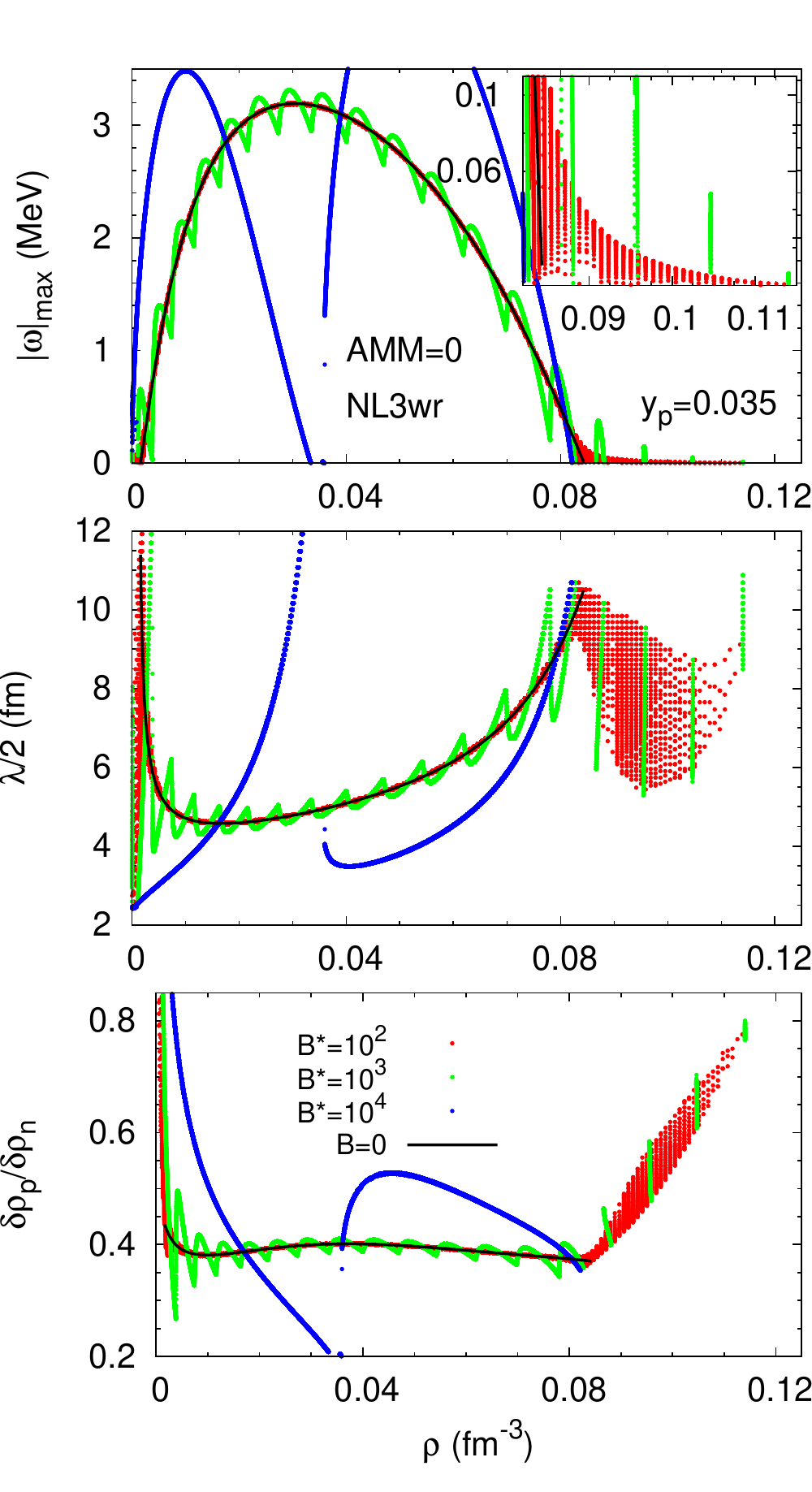} &
\includegraphics[width=0.5\textwidth]{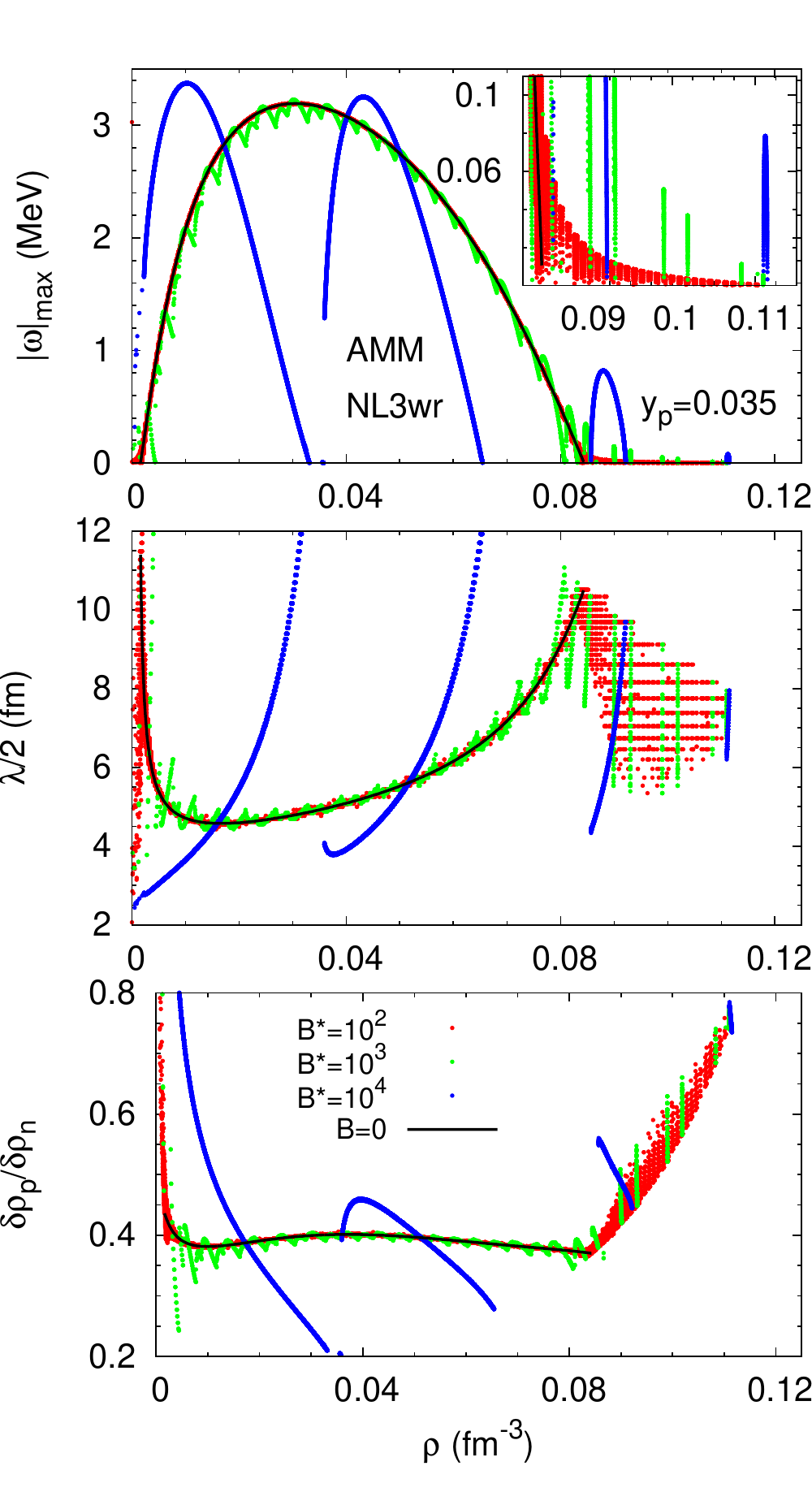} \\
\end{tabular}
\caption{Largest growth rate (top panels), the corresponding half-wavelength (middle panels), and the proton-neutron density fluctuation ration (bottom panels) versus density for NL3$\omega\rho$, with a proton fraction of $y_p=0.035$, and without (left panels) and with (right panels) AMM. Three different values of $B$ are considered: $B^*=10^2$ (red), $B^*=10^3$ (green), and $B^*=10^4$ (blue). A comparison with the $B=0$ (black lines) results is also made. }
\label{fig5}
\end{figure*}

In the following, we shall consider that the behavior of the system 
will be determined by the largest growth rate, $|\omega|_{max}$, the mode that drives the system into a separation of a high  and a low density phases. As in Ref. \cite{providencia06}, we will
consider that half wave-length of the maximum growth rate mode is a good estimation of the order of magnitude of the size of the clusters which are formed,  as shown in Ref. \cite{avancini08}, where the size of the clusters 
obtained  within a  Thomas-Fermi calculation were compared with the half-wave length
associated to the most unstable mode. These quantities are plotted in Figs. \ref{fig4}, and \ref{fig5}. \\

In Fig. \ref{fig4}, the largest growth rates (top panels), the
corresponding half-wave length (middle panels), and the
$\delta\rho_p/\delta\rho_n$ ratio are shown for $B^*=10^3$.  The left panels show the calculations performed without AMM, and the right panels take the AMM into account. In all figures,
the $B=0$ results are represented by  a black curve, in grey we
  show the results with a fixed proton fraction of 0.035, and in blue
  we take, above the $B=0$ crust-core transition density, the proton
  fraction as the one found in $\beta-$equilibrium matter.  {In
    Fig. \ref{fig5}, the same quantities calculated with  three
    different magnetic field intensities, $B^*=10^2,
    10^3$, and $10^4$, are compared. In this figure, all results were  obtained with a fixed proton fraction $y_p= 0.035$.}

{\small

\begin{table}[!htbp]
\caption{Transition densities and pressures for the magnetic fields considered in this study, together with the correspondent fractional moment of inertia of the neutron star crust, for a star of $M=1.4$ M$_\odot$ and $R=13.734$ km. Also shown are the crust thicknesses, $\Delta R$, the thickness due to the inhomogeneous region found when $B\neq$0, $\Delta R'=R(\rho_1)-R(\rho_2)$, and the difference to the $B=0$ result, $\Delta R_B=\Delta R-\Delta R(B=0)$. The results shown take into account AMM. The transition densities are calculated when $|\omega|=0$ (see top panels of Figs. \ref{fig4} and \ref{fig5}, and the text for more details). The values for $\rho_2$ correspond to the calculations with a  $\beta-$equilibrium matter proton fraction.  }  
\label{tab1}
  \begin{tabular}{ c cccc cccc cccc}
    \hline
    \hline
  $B^*$ &   $\rho_1$   &  $\rho_2$ 
   & $P_1$  &   $P_2$ &$\Delta R$ & $\Delta R'$  &
   $\Delta R_B$ & $\frac{\Delta I_{cr}}{I}$ \\ 
    &   (fm$^{-3}$)  &   (fm$^{-3}$) & 
 $\left(\frac{\mbox{MeV}}{\mbox{fm}^3}\right)$ &  $\left(\frac{\mbox{MeV}}{\mbox{fm}^3}\right)$ & (m)
    & (m) & (m)\\

    \hline
0 &  0.0843 &  0.0843  & 0.5196 &   0.5196 & 1368 & 0 & 0  &  0.0676\\
$10^2$ &   0.0837   & 0.1044  & 0.5119  &  0.8541  & 1551 & 185 & 182  &  0.0968 \\      
  $10^3$  &  0.0808   &  0.1096   &  0.4758  &  0.9743  & 1609 & 257 & 240 & 0.1056 \\ 
  $10^4$  &   0.0654   &  0.0998   & 0.3274  &  0.8095  &  1503 & 260  &  134 &  0.0922\\ 
    \hline
    \hline
  \end{tabular}
\end{table}
}

The effect that had
already been identified with the spinodal sections is clearly shown in
theses figures: the unstable regions occur well beyond the $B=0$ 
crust-core transition at 0.0843 fm$^{-3}$, and extend until $\rho_t=0.11$~fm$^{-3}$, 40\% larger than
the $B=0$ transition density. However, above 0.082 fm$^{-3}$, the
unstable regions alternate with stable ones. This effect was first
discussed in Ref. \cite{fang16}. The transition densities that define
  the limits of the region of alternating stable and unstable regions have been labelled
  $\rho_1$ and $\rho_2$, and are shown in Table \ref{tab1}. The transition
  density $\rho_1$ defines  the first time $|\omega|$ goes to zero, and
  the density $\rho_2$
  defines the onset of the homogeneous matter, meaning that we have a
  range of densities between $\rho_1$
  and $\rho_2$ where unstable and stable regions alternate. At $B=0$, both
  densities coincide, i.e.  $\rho_1=\rho_2$.
The density $\rho_2$ is determined by taking the  $\beta-$equilibrium matter
  proton fraction above the $B=0$ crust-core transition, and $\rho_1$
  is calculated with the fixed $y_p=0.035$ proton fraction, obtained
  from the $B=0$ calculations of the pasta phases,  since
  $\rho_1$ is a  density that lies below the $B=0$ crust-core transition. Taking the
  $\beta$-equilibrium proton fraction, instead of the fixed 0.035, has a
  non-negligible effect, and, in fact, reduces the instability region,
  because the $\beta$-equilibrium condition predicts larger proton
  fractions, and the larger the proton densities, the smaller the
  effects due to the magnetic field. { In particular, for
    $B^*=10^4$, $\rho_2$ is $\sim 0.01$ fm$^{-3}$ smaller, taking
    $y_p^{\beta-eq}$ instead of $y_p=0.035$. 
This difference is $\sim 0.005$ fm$^{-3}$ for $B^*=10^2$, and $\sim 0.002$ fm$^{-3}$ for $B^*=10^3$.
The discrete feature of the
  Landau levels results in a non-monotonic behavior of this quantity
  for the larger values of $B^*$.}

 Besides  $\rho_1$ and $\rho_2$, in
  Table \ref{tab1}, we also give 
the pressure at these two densities and the fractional
  moment of inertia of the crust,  a quantity that depends directly on
  the pressure and density at the crust-core transition, and that has an
  important impact in explaining pulsar glitches.  The fractional
    moments of inertia  were
    calculated from the approximate expression given in \cite{Worley08,Lattimer00}:
\begin{eqnarray}
\frac{\Delta I_{cr}}{I}&\simeq& \frac{28\pi P_t R^3}{3M}\frac{(1-1.67\beta-0.6\beta^2)}{\beta} \nonumber \\  &\times&\left[1+\frac{2P_t(1+5\beta-14\beta^2)}{\rho_t m_b\beta^2}\right]^{-1} \, , 
\end{eqnarray}
where $\Delta I_{cr}$ is the crust moment of inertia, $I$ is the
total star moment of inertia, $P_t$ and  $\rho_t$ are the crust-core transition
pressure and density, respectively, $M$ and $R$ are the gravitational mass and
radius of the star, $\beta=GM/R$ is the compactness parameter, and
$m_b$ is the nucleon mass. In Table \ref{tab1}, the crust thickness,
$\Delta R=R(0)-R(\rho_2)$, the thickness of the region
between $\rho_1$ and $\rho_2$, $\Delta R'=R(\rho_1)-R(\rho_2)$, and
the difference between the crust thicknesses at $B=0$ and $B\ne 0$, $\Delta R_B=\Delta R-\Delta R(B=0)$, are also
displayed. These results take into account the  AMM, and they have
been calculated  for a star with $M=1.4$ M$_\odot$, and a radius of
$R=13.734$ km. For the calculation of the fractional moment of inertia
of the crust, we took for $P_t$ and $\rho_t$ the values of $P_2$ and
$\rho_2$, given in the Table for each magnetic field. Our results for $B=0$ agree with  the transition
  densities and pressures and the moment of inertia of the
  crust obtained  in Ref. \cite{fattoyev10}, see tables III and IV. This is expectable, since  the
 same expression for the crustal moment of inertia has been used,  as
 in Refs. \cite{Worley08,Lattimer00}.

  The magnetic field gives rise to  larger values of the crust-core transition pressure and density, and these affect
directly $\Delta I_{crust}/{I}$. These values are 
  much higher than the prediction in \cite{link99,glitch2} for the
  Vela pulsar, { 0.016, when no entrainment effects are considered,}
 and they would be high enough for the crust to completely describe
 the  glitch mechanism, even taking into account the effect of
 entrainment  \cite{glitch2,chamel13}.  { In fact, in this case the ``effective''
   moment of inertia associated with the fluid is lowered  and the
   constraint inferred from glitches requires that the crustal moment
   of inertia is $\langle m^*_n \rangle/m_n\sim 4-6$ larger
   \cite{glitch2}, where $ m^*_n$ is the effective neutron mass including
 entrainment, and $m_n$ the bare neutron mass.
   To explain  the Vela glitches,  this constraint would be equivalent
   to requiring a fractional  crustal moment of inertia $\sim
   0.064-0.096$. 
}

The calculation including the AMM of protons and neutrons presents
twice as much unstable regions due to the separation of each proton
Landau level in two,  with a different spin
polarization. The resulting regions are narrower and have  smaller  growth rates. 
More information on the properties of this range of
densities is obtained from the middle and bottom panels of both Figures. In the middle
panel, the half-wave length of the perturbation is plotted, and it gives an
estimation of the size of the cluster that will be formed. Within each
of these independent unstable regions, the cluster size changes from
about 9 fm to about 4 fm in a very narrow density range. Finally, the bottom
panel gives some information on the proton content of the dense phase:
the clusters will be quite proton rich with a proton-neutron density
fluctuation ratio well above the 0.04 ratio of the homogeneous matter.

\begin{figure}[!htbp]
\includegraphics[width=0.45\textwidth]{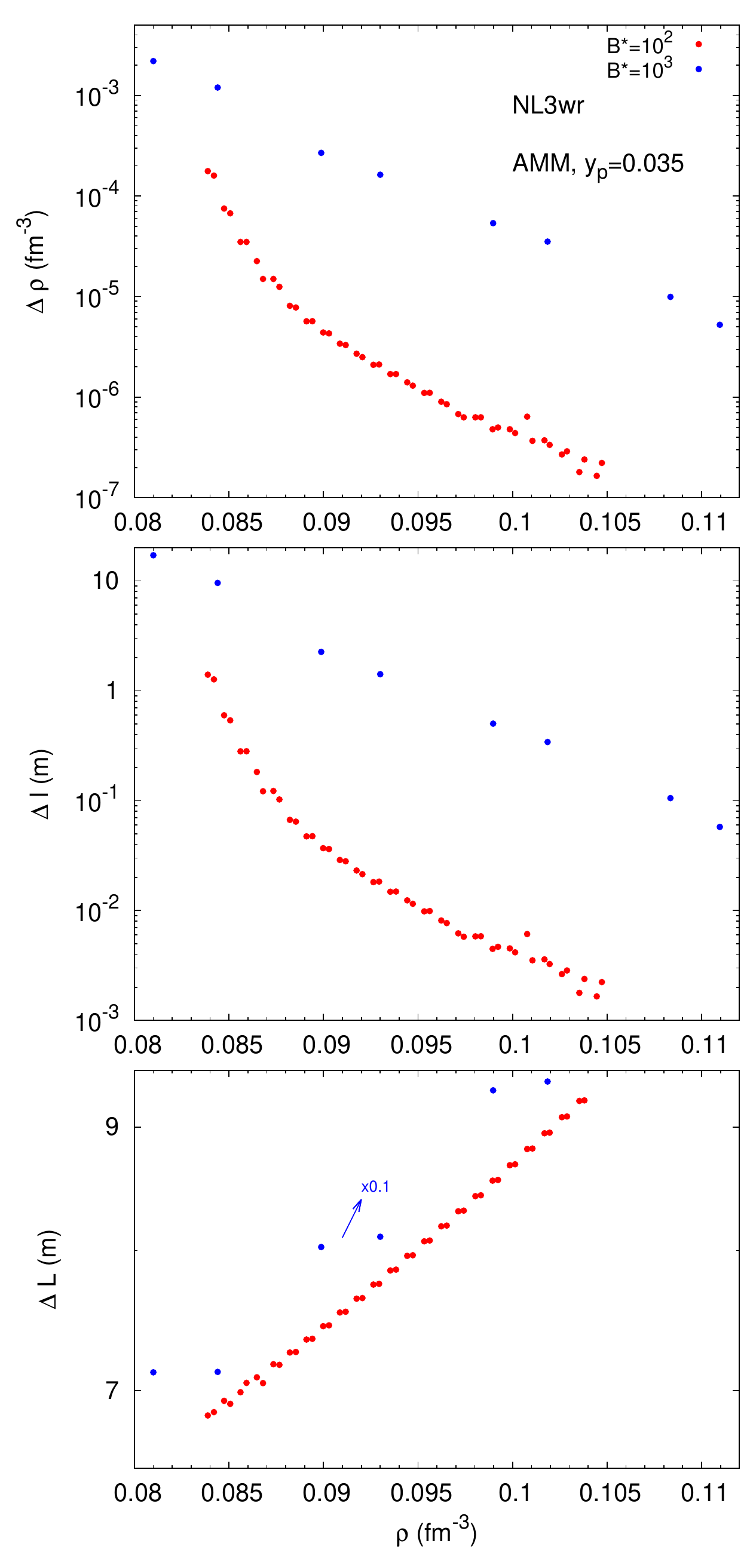} \\
\caption{Width of the instability peaks, $\Delta \rho$, (top panel),
  the thickness of the instability regions, $\Delta l$
  (middle panel), and the distance between the instabilities, $\Delta
  L$ (bottom panel), as a function of the baryonic
  density, for NL3$\omega\rho$, a proton fraction of $y_p=0.035$,
  $B^*=10^2$ (red), and $B^*=10^3$ (blue), with AMM. In the bottom
  panel, $\Delta L$ obtained with   $B^*=10^3$ is multiplied by a factor of 0.1.}
\label{fig6}
\end{figure}

For  $B^*=10^2$ and $10^3$, we have made an analysis of the dependence
on the density of the width and
separation  of the instability peaks above the transition density,
$\rho_t(B=0)$. In the top panel of Fig. \ref{fig6}, the
width of these peaks, $\Delta \rho$, is plotted versus the density, and it is seen
that it  decreases exponentially, showing a steeper decrease for
$B^*=10^2$. We may ask what is the impact of these structures in the
neutron star, and, in particular, what is their size. In order to make
an estimation, we have considered a 1.4 $M_\odot$ star profile, and
calculated the localization of the instabilities. This  has allowed
the  determination of the thickness of the instability regions, $\Delta l$, in
meters, and the distance between the instabilities, $\Delta L$, also in
meters. The results are shown in the middle and bottom panels of
Fig. \ref{fig6}, respectively. Some conclusions are in order: for
$B^*=10^2$,  the spacing between peaks increases with density from
$\sim 6$ m to $~\sim 9$ m, until the homogeneous core sets in, while
for $B^*=10^3$, this spacing is one order of magnitude larger; the
width of the peaks decreases exponentially from $\sim 1$ m for
$\rho=0.085$fm$^{-3}$ to $\sim 5$ mm for $\rho=0.1$fm$^{-3}$ for
$B^*=10^2$, and is 1-2 orders of magnitude larger if $B^*=10^3$. In the bottom panel, we may identify two almost parallel curves which correspond to the distance between peaks with the polarization aligned or anti-aligned with the magnetic field.

These properties indicate that at the crust-core transition, matter is very
complex and that the magnetic field favors the  large charge
concentration in the clusters. Even considering matter below the
$B=0$ crust-core transition, the present calculation indicates that 
there is a fast change of the cluster size. This change will probably
cause the cluster size and structure to change more strongly with density
than it would be expected for $B=0$, giving rise to more 
heterogeneous matter.

Simulations  of the
time evolution of the magnetic field at the crust have shown that 
the existence of  amorphous and heterogeneous  matter deep in
the inner crust, with a high impurity parameter and, therefore, highly resistive,
favors a fast decay of the magnetic fields. This has been proposed as
an explanation  for the non-observation of
 x-ray pulsars with a  period above 12~s \cite{pons13}.

\section{Conclusions}

We present a formalism to calculate the
eigenmodes of nuclear and stellar matter in the presence of very strong magnetic fields, as may be found in
magnetars. The AMM of protons and neutrons is taken into account and it is 
shown that it is not negligible for magnetic fields above 10$^{15}$ G.
 Within this formalism, it is possible to estimate the crust-core  transition inside
neutron stars from the spinodal section, the surface where the
eigenmodes go to zero. Inside these surfaces, matter will separate into
a dense and a gas phase. This RMF description of stellar matter takes
into account both the Coulomb field and finite size effects related to
the finite range of nuclear force. We have only considered the propagation of waves in the
direction of the magnetic field.

The main objective of this work was to complete the discussion
published in Ref. \cite{fang16}, where the linearized Vlasov equation was used to
determine the importance of taking into account explicitly the
effect of magnetic fields of the order of 10$^{15}$-$10^{17}$ G on the
description of stellar matter at subsaturation densities, in
particular, the effect of the magnetic field on the crust-core
transition in magnetized neutron stars.
 It was shown in Refs. \cite{broderick,broderick02} that only
fields two to three orders of magnitude stronger than the critical
electron  magnetic field 
have a finite effect on
the EoS at densities above saturation density, but no discussion has
been presented for lower densities of interest for the inner crust.
However, in
Refs. \cite{chamel12,chamel15}, the non-negligible effect of the magnetic field on the
properties of the outer crust has already been discussed. 
 Even though the largest fields
detected on the surface of magnetars are 2$\times 10^{15}$ G on the
SGR 1806−20 (see e.g. \cite{olausen14} and the McGill Online Magnetar
Catalog \cite{sgr}), we expect that stronger fields  are still realistic, because in Refs. \cite{kiuchi08,rezzolla12} equilibrium configurations
of magnetars with fields well above 10$^{15}$ G in the interior have
been obtained when a toroidal component of the magnetic field is  considered.

The effect of the Landau quantization of the levels of charged particles  gives
rise to a spinodal section that presents a structure of  bands at the border between
clusterized and homogeneous matter. As a consequence, it was shown
that the transition between the crust and the core of magnetars is defined by a complex region that is  $\sim 0.02-0.04$ fm$^{-3}$ wide,
characterized by a succession of  homogeneous and  clusterized matter.
Calculations have been performed with and without the inclusion of the
AMM, and it was shown that the AMM has a non-negligible effect, and,
moreover, it contributes to an extra complexity of matter at the
crust-core transition.

The determination of the mode associated with the maximum growth rate
 in the density range delimited by the spinodal section, allowed the
 estimation  of several properties of the
 clusters that are formed in the unstable regions, in particular, the
 size and a qualitative charge content. Two different  density regions
 should be considered independently: a) inside the density region
 delimited by the $B=0$ spinodal, the size and the $\delta
 \rho_p/\delta\rho_n$  ratio change along the trend defined by the
 corresponding $B=0$ results, showing fluctuations around these
 values, the larger the field the larger the fluctuations; 
b) the density region outside the $B=0$ spinodal section, where  an
 alternation of unstable and stable regions occurs. In this region, the
 clusters formed inside the unstable regions vary between $\sim 5-10$
 fm inside very small density ranges, and the larger the densities,
 the larger the proton fraction inside the clusters.
In a 1.4 $M_\odot$ star, this region has a width of $\sim 200$ m for
 $B^*=10^2$, and a bit larger for $B^*=10^3$ and $10^4$. If $B^*=10^2$,
 the spacing between peaks  is $\sim 6-9$ m, and
 the width of the peaks decreases exponentially from 1 m at the $B=0$
 crust-core transition,  to  $2\times 10^{-3}$ m at the onset of
 homogeneous matter.  {For $B^*=10^3$ these quantities are 1-2 orders of
 magnitude larger.} Including this  complex region in the crust, the
crust moment of inertia can be as large as 9-10\% the total star
moment of inertia, circa 30\% larger than the ratio obtained for
$B=0$.

These results indicate that it is necessary to study the transport
properties, such as electric conductivity and shear viscosity, of this
complex matter, see \cite{yakovlev15,yakovlev15a}.
 Also,  the overall properties
obtained for stellar 
matter at subsaturation densities seem to support the occurrence of
a  larger electrical  resistivity  in the presence of a strong magnetic
field,  supporting the existence of a resistive layer
deep inside the inner crust of magnetized neutron stars, as proposed
in Ref. \cite{pons13}, that would cause a fast decay of the magnetic field, and explain the
non-observation of isolated pulsars,  with periods larger than 12 s.

\section*{ACKNOWLEDGMENTS}

This work  is  partly  supported  by  
the FCT (Portugal) project UID/FIS/04564/2016, 
and by ``NewCompStar'', COST Action MP1304. H.P. is supported by FCT (Portugal) under Project No. SFRH/BPD/95566/2013, and J.L. is supported by National Natural Science Foundation of China (Grant No.
11604179), and Shandong Natural Science Foundation (Grant No. ZR2016AQ18).

\section*{Appendix}

The coefficients $a_{ij}$ of the matrix (\ref{det}) can be written as:

\begin{eqnarray*}
a_{11}&=&\sum_{\nu,s}\frac{g_s}{2\pi^2}\frac{G_{\phi_p}M^*}{\omega_s^2}\frac{eB}{P_F^p(s_p^2-1)},\\
a_{12}&=&-\sum_{\nu,s}\frac{eB}{2\pi^2 V_F^p(s_p^2-1)}\left[\left(1-\frac{\omega^2}{k_z^2}C_{1p}\right)g_v D_{1p} \right. \nonumber \\
&+&\left.\left(1-\frac{\omega^2}{k_z^2}C_{2p}\right)\frac{g_\rho}{2}D_{2p}
+\left(1-\frac{\omega^2}{k_z^2}\right)\frac{e^2}{\omega_A^2}\right]
-1,\\
a_{13}&=&\sum_{\nu,s}\frac{g_s}{2\pi^2}\frac{G_{\phi_p}M^*}{\omega_s^2}\frac{1}{V_F^p\epsilon_F^n (s_p^2-1)},\\
a_{14}&=&-\sum_{\nu,s} \frac{1}{2\pi^2 V_F^p(s_p^2-1)} \left[ \left(1-\frac{\omega^2}{k_z^2} C_{1p}\right)g_v D_{1n} \right.\nonumber \\
&+& \left. \left(1-\frac{\omega^2}{k_z^2} C_{2p}\right)\frac{g_\rho}{2}D_{2n}\right], \\
a_{15}&=&\sum_{\nu,s}\left(1-\frac{\omega^2}{k_z^2}\right)\frac{e^2}{2\pi^2\omega_A^2}
\frac{eB}{V_F^p(s_p^2-1)} ,
\end{eqnarray*}

\begin{eqnarray*}
a_{21}&=&\frac{\bar m_p}{\tilde m_p}a_{11}-1, \quad 
a_{22}=\frac{\bar m_p}{\tilde m_p}(a_{12}+1),\\
a_{23}&=&\frac{\bar m_p}{\tilde m_p}a_{13}, \quad 
a_{24}=\frac{\bar m_p}{\tilde m_p}a_{14}, \quad
a_{25}=\frac{\bar m_p}{\tilde m_p}a_{15} ,
\end{eqnarray*}

\begin{eqnarray*}
a_{31}&=&\sum_{s}\frac{g_sM^*}{(2\pi)^2}\frac{eB}{\omega_s^2\epsilon_F^p}\epsilon_F^n P_F^n L^*(s_n),\\
a_{32}&=&-\sum_{s}\left[\left(1-\frac{\omega^2}{k_z^2}C_{1n}\right)g_v D_{1p} 
-\left(1-\frac{\omega^2}{k_z^2}C_{2n}\right)\frac{g_\rho}{2}D_{2p}\right]  \nonumber \\
&&\times\frac{eB}{4\pi^2}\epsilon_F^n P_F^n L(s_n), \\
a_{33}&=&\sum_{s}\frac{g_sM^*}{(2\pi)^2}\frac{P_F^n}{\omega_s^2} L^*(s_n),\\
a_{34}&=&-\sum_{s}\left[\left(1-\frac{\omega^2}{k_z^2}C_{1n}\right)g_v D_{1n}- 
\left(1-\frac{\omega^2}{k_z^2}C_{2n}\right)\frac{g_\rho}{2}D_{2n}\right] \nonumber \\
&&\times\frac{\epsilon_F^n P_F^n L(s_n)}{4\pi^2} -1,\\
a_{35}&=&0, 
\end{eqnarray*}

\begin{eqnarray*}
a_{41}&=&a_{31}\frac{H^*(s_n)}{L^*(s_n)}, \quad
a_{42}=a_{32}\frac{H(s_n)}{L(s_n)},\\
a_{43}&=&a_{33}\frac{H^*(s_n)}{L^*(s_n)}-1, \\
a_{44}&=&(a_{34}+1)\frac{H(s_n)}{L(s_n)}, \quad
a_{45}=0,
\end{eqnarray*}

\begin{eqnarray*}
a_{51}&=&a_{53}=a_{54}=0,\\
a_{52}&=&\sum_{\nu,s}\left(1-\frac{\omega^2}{k_z^2}\right)\frac{e^2}{2\pi^2\omega_A^2}
\frac{eB}{V_F^e(s_e^2-1)}, \\
a_{55}&=&-\sum_{\nu,s}\left(1-\frac{\omega^2}{k_z^2}\right)\frac{e^2}{2\pi^2\omega_A^2}
\frac{eB}{V_F^e(s_e^2-1)}-1 \\
\end{eqnarray*}
where $\tilde m_p=\bar m_p+s\mu_{N}\kappa_{p}B $, $\omega_s^2=\omega^2-k_z^2-m^2_{s,eff}$, $\omega_A^2=\omega^2-k_z^2$,  and $s_i=\frac{\omega}{\omega_{0i}},\,i=p,n,e.$

The remaining coefficients are given by
\begin{eqnarray*}
G_{\phi_p}&=&\frac{g_sM^*\bar m_p}{\epsilon_F^p(\bar m_p+s\mu_{N}\kappa_{p}B)},\nonumber \\
G_{\phi_n}&=&\frac{g_sM^*}{\epsilon_F^n}\left(1-\frac{s\mu_N\kappa_n B}{\sqrt{M^{* 2}+P_F^{n2}(1-x^2)}}\right), \nonumber 
\end{eqnarray*}
\begin{eqnarray*}
 C_{1i}&=&1-\frac{\tau_i g_\rho\chi}{2g_v\omega_\rho^2}, \\
C_{2i}&=&1-\frac{2\tau_i g_v\chi}{g_\rho\omega_v^2}, \nonumber 
\end{eqnarray*}
\begin{eqnarray*}
D_{1i}&=&\frac{\tau_i g_\rho\chi}{2
          D_\nu}+\frac{g_v\omega_\rho^2}{D_\nu}, \\
 D_{2i}&=&\frac{g_v\chi}{D_\nu}+\frac{\tau_i g_\rho\omega_v^2}{2D_\nu}, \nonumber \\
D_\nu&=&\omega_v^2\omega_\rho^2-\chi ,\nonumber 
\end{eqnarray*}
\begin{eqnarray*}
 L(s_n)&=&\int _{-1}^{1}\frac {x}{(s_n-x)}dx, \quad L^*(s_n)=\int _{-1}^{1}\frac {G_{\phi_n}xdx}{(s_n-x)}, \nonumber \\
H(s_n)&=&\int _{-1}^{1}\left(1-\frac{s\mu_N\kappa_n B}{\sqrt{M^{* 2}_n+P_F^{n2}(1-x^2)}}\right)\frac {xdx}{(s_n-x)}, \nonumber \\
H^*(s_n)&=&\int _{-1}^{1}\left(1-\frac{s\mu_N\kappa_n B}{\sqrt{M^{* 2}_n+P_F^{n2}(1-x^2)}}\right)\frac {G_{\phi_n}xdx}{(s_n-x)}. \nonumber
\end{eqnarray*}

The amplitudes of the scalar densities fluctuations, $A^{is}_{\omega,\nu,s}$, and of the vector densities fluctuations, $A^{i}_{\omega,\nu,s}$, are given by
\begin{eqnarray*}
A_{\omega,\nu,s}^{ps}&=&\sum_\xi\xi\frac{\bar m_p}{(\bar m_p+s\mu_{N}\kappa_{p}B )}S_{\omega,\nu,s}^p(\xi),\\
A_{\omega,s}^{ns}&=&\int _{-1}^{1}P_F^{n2}x\left(1-\frac{s\mu_N\kappa_n B}{\sqrt{M^{* 2}_n+P_F^{n2}(1-x^2)}}\right)S_{\omega,s}^{n}dx,\\
A_{\omega,\nu,s}^p&=&\sum_\xi\xi S_{\omega,\nu,s}^p(\xi),\\
A_{\omega,s}^{n}&=&\int _{-1}^{1}P_F^{n2}xS_{\omega,s}^{n}dx,\\
A_{\omega,\nu,s}^e&=&\sum_\xi\xi S_{\omega,\nu,s}^e(\xi),\\
\end{eqnarray*}

\end{document}